\documentstyle[11pt]{article}
\begin{document}
\begin{flushright}
\vspace{-3.0ex} 
    {ADP-99-41/T378} \\
\vspace{-2.0mm}
\vspace{3.0ex}
\end{flushright}
\centerline{\bf\large SPECTRA OF BARYONS CONTAINING TWO HEAVY QUARKS}
\vspace{0.3cm}
\centerline{\bf \large IN POTENTIAL MODEL}

\vspace{1cm}

\centerline{\small Sheng-Ping Tong$^1$, Yi-Bing Ding$^{1,2}$, Xin-Heng Guo$^{3,4}$,
Hong-Ying Jin$^{5}$,}

\vspace{0.2cm}
\centerline{\small Xue-Qian Li$^{6,7}$, Peng-Nian Shen$^{4,6}$ and
Rui Zhang$^{7}$}

\vspace{0.8cm}
{\small

1. Graduate School, USTC at Beijing, Academia Sinica, P.O. Box 3908, Beijing,
100039, China\\

2. Department of Physics, University of Milan, INFN, Via Celoria 16, 20133
Milan, Italy\\

3. Department of Physics and Mathematical Physics, and Special
Research Center for the Subatomic Structure of Matter, University
of Adelaide, SA 5005, Australia\\

4. Institute of High Energy Physics, Academia Sinica, P.O. Box 918-4, Beijing
100039, China\\

5. Department of Physics, University of Mainz, Germany\\

6. CCAST(World Laboratory), P.O. Box 8730, Beijing 100080, China\\

7. Department of Physics, Nankai University, Tianjin 300071, China}

\vspace{1cm}

\begin{center}
\begin{minipage}{12cm}
\noindent{\bf Abstract}

In this work, we employ the effective vertices for interaction between
diquarks (scalar or
axial-vector) and gluon where the form factors are derived in terms of the B-S
equation, to obtain the potential for baryons including a light quark and a heavy
diquark. The concerned phenomenological parameters are obtained by fitting
data of $B^{(*)}-$mesons instead of the heavy quarkonia. The operator ordering
problem in quantum mechanics is discussed. Our numerical results indicate
that the mass splitting between $B_{3/2}(V),B_{1/2}(V)$ and $B_{1/2}(S)$ is
very small and it is consistent with the heavy quark effective theory (HQET).
\end{minipage}
\end{center}

\vspace{2cm}

\baselineskip 22pt

\noindent{\bf I. Introduction}

It is well known that the heavy flavor physics can be different from the world
where only light flavors are involved in many aspects. Since the heavy
flavor can serve as  a static color source as $M_Q\gg \Lambda_{QCD}$, an extra
symmetry $SU_f(2)\otimes SU_s(2)$ exists \cite{Isgur} which can attribute
all non-perturbative QCD effects into one or a few form factors and make the
hadronic matrix elements evaluation much simplified. On the other hand, the
diquark structure
in baryons causes interests of many theorists of high energy physics
\cite{Anse1}. The
possible diquark structure in nucleons has been studied in the non-relativistic
QCD-based potential quark model \cite{Fred}. Even though as pointed out,
a spin-0 diquark structure may exist in nucleons, one has reason to doubt its
validity. Because the two light quark which are supposed to constitute a
diquark are relativistic and dispersive in space, it is not very likely to
compose a tight object. To compensate the spatial dispersion, one
can introduce some form factors at the effective vertices \cite{Anse}.

On contraries, we can be convinced that if there are two heavy quarks ($bb,bc,
cc$) in the baryons, they would tend to constitute a substantial diquark with
small spatial dispersion which can serve as a static $\bar 3$ color source
for the light quark \cite{Falk}\cite{Guo}.
Savage and Wise estimated spectra of baryons
with two heavy quarks in the heavy quark effective theory (HQET) \cite{Savage}.
Recently, Ebert et al. evaluated the spectra of such baryons in terms of the
local Schr\"{o}dinger-like quasipotential equation \cite{Ebert}.
In the framework of the potential model, the interaction between
the light quark and heavy diquark can be derived
by calculating their elastic scattering amplitudes \cite {Landau}, but
the key point is the form of the effective vertices at the diquark-gluon
interaction. Similarly, Gershtein et al. also considered the spectroscopy of
doubly charmed baryons where they include angular and radial excited states
\cite{Ger}.

As aforementioned, even though the diquark consisting of two heavy quarks is
tight, in reality it is still not point-like and the
small deviation should be estimated. The authors of ref.\cite{Ebert} introduced
simplified form factors and ignored their $k^2-$dependence to take the effects
into account.

In this work, we re-derive the effective potential by using the B-S equation
and obtain the effective vertices. In the derivations, the $k^2-$dependence
is retained explicitly. We find that this dependence leads to an extra
Yukawa-type term.

Because of the serious relativistic effects of the light quark, the non-relativistic
expansion of the potential becomes dubious, since we must truncate the expansion,
usually to order ${\bf p}^{\; 2}/m^2$, where $m^2$ may be the light quark mass.
Generally, ${\bf p}^{\; 2}\sim \Lambda_{QCD}^2\sim (0.2\; GeV)^2$, $m$ takes the
constituent quark mass as $m_u\approx m_d\sim 0.33$ GeV, thus the expansion
factor ${\bf v}^{\; 2}/c^2={\bf p}^{\; 2}/m^2\sim 0.37$, so the next leading order should be
of certain contributions. In fact, one can attribute such uncertainties to the
parameters which exist in the potential, but cannot be directly measured.
Therefore considering this situation,  instead
of the $\Upsilon$ and $J/\psi$ spectra we re-fit the data of $B^{(*)}-$meson,
which includes a light quark and
a heavy b-quark, to obtain the parameter values. Substituting the re-obtained
parameters into the potential we evaluate the baryon spectra, one has reason to
expect that in this way, the errors can be substantially reduced.

In the process, we also consider the operator-ordering problem. Since we derive
the scattering amplitude in the momentum space, all quantities are commutative.
But when the Fourier transformation with respect to the exchanged momentum
$\bf k$ is carried out, the coordinate $\bf r$ and momentum $\bf p$
co-exist, their ordering becomes a problem. This is an inconsistency which
exists in the treatment. However, our results (see below) indicate that
the ordering only determines the parameter values, but the final measurable
quantities, i.e. the spectra do not deviate much from each other in different
ordering schemes.

We also briefly discuss the possible mixing problem and point out the drawbacks
of the potential model.

The paper is organized as following. After this introduction, we present the
formulation and explicit form of the potential. We also discuss some concerned
problems such as the ordering schemes. In Sec.III, we discuss how to obtain
the spectra in terms of variational method, trial function and concerned
issues. In Sec.IV, we give the numerical results and the adopted parameters.
The last section is devoted to our conclusion and discussion.\\

\noindent{\bf II. Formulation}

(a) The effective vertices for diquark-gluon coupling.

Since the diquark is not rigorously a point-like subject, we cannot simply
use the vertices in the fundamental QCD theory. Instead, we derive such
effective vertices with certain form factors in terms of the B-S equation.
The diquark contains two heavy quarks which constitute a color $\bar 3$ triplet,
for this bound state the Cornell potential would be a good approximation
and we use it as the B-S kernel.

We derive the effective vertices for $SSg,AAg$ and $SAg,ASg$ as following:
\begin{equation}
\langle  S^{\prime}(v^\prime)|J_\mu |S(v)\rangle=\sqrt{MM^\prime}
(f_1v^\prime _\mu +f_2v_\mu)\;\;\;\;{\rm for\; SSg\; coupling},
\end{equation}
\begin{eqnarray}
\label{AAJ}
\langle  A^{\prime}(v^\prime ,\eta^\prime)|J_\mu|A(v,\eta)\rangle
&=& \sqrt{MM^\prime}[f_3(\eta\cdot\eta^{\prime\ast})v^\prime_\mu
+f_4(\eta^{\prime\ast}\cdot\eta)v_\mu
\nonumber \\
&+& f_5(\eta\cdot v^\prime)(\eta ^{\prime\ast}\cdot v)v^\prime_\mu
+f_6(\eta\cdot v^\prime)(\eta ^{\prime\ast}\cdot v)v_\mu
\nonumber \\
&+& f_7\eta^{\prime\ast}_\mu (\eta\cdot v^\prime)
+f_8(\eta^{\prime\ast}\cdot v)\eta _\mu
\nonumber \\
&+& f_9i\epsilon _{\mu\nu\rho\sigma}\eta^{\prime\ast\nu}\eta^\rho
v^{\prime\sigma} \nonumber \\
&+& f_{10}i\epsilon _{\mu\nu\rho\sigma}\eta^{\prime\ast\nu}\eta
^\rho v^\sigma] \;\;\;\;{\rm for\; AAg\; coupling},
\end{eqnarray}
\begin{eqnarray}
\label{SAJ}
\langle  A^{\prime}(v^\prime,\eta^\prime)|J_\mu|S(v)\rangle
&=& \sqrt{MM^\prime}[f_{11}\eta^{\prime\ast}_\mu
+f_{12}(\eta^{\prime\ast}\cdot v)v^\prime_\mu
\nonumber \\
&+& f_{13}(\eta^{\prime\ast}\cdot v)v_\mu
+f_{14}i\epsilon _{\mu\nu\rho\sigma}\eta^{\prime\ast\nu}v^{\prime\rho}v^\sigma]
\;\;\;\;{\rm for\; ASg\; coupling},
\end{eqnarray}
where $S$ and $A$ stand for scalar and axial-vector diquarks,
$v^{\prime},v,\eta^{\prime},\eta, M^{\prime}, M$ are the four-velocities,
polarization vectors (for axial-vector diqaurk only), and masses of
the diqaurks in the
"final" and "initail"states of the scattering respectively. The
corresponding form factors are derived by solving the transition B-S equation
and the details were given in our previous work \cite{Guo}. In our case, we
find relations
\begin{equation}
f_1=f_2=f_7=f_8=-f_3=-f_4=f_{14}=f,
\end{equation}
and
\begin{equation}
\label{par}
f_9=f_{10}=f_{11}=f_{12}=f_{13}=0,
\end{equation}
and the relations (\ref{par}) can be realized by simple parity analysis.
Obviously, the terms related to $f_5$ and $f_6$ are proportional to $|{\bf v}|^3
$ ($|{\bf p}|^3/m^3$) so that can be neglected as we only keep the
non-relativistic expansion up to order of ${\bf p}^{\; 2}$.

Here we would like to draw attention of readers that in expressions
(\ref{AAJ}) and (\ref{SAJ}), the order of $\eta$ and $\eta^{\prime *}$ is
not trivial. When we derive these formulae in Quantum Field Theory (QFT), they are commutative, so
that we can put them in any order, however, as we turn $\eta$ and $\eta^{\prime*}$
into spin-operators of Quantum Mechanics (QM), the order problem emerges. Because  
${\bf S}-$operator
is not self-commutative, thus in the operator form, $\eta\cdot\eta^{\prime *}
\neq \eta^{\prime *}\cdot \eta$. Therefore, when we write the expressions,
we must be very careful about the order. Our strategy is that we keep the right
forms of the leading terms relating to spins such as the spin-orbit coupling
terms to choose the appropriate
orders. In fact, we can also determine the order by analyzing the symmetry
of the whole Hamiltonian as we have done in the expressions (\ref{AAJ}) and
(\ref{SAJ}).

The form factors $f_i$ involves the B-S integrals and cannot be analytically
expressed. One can only obtain the numerical results instead. However, in order
to serve our final goal to derive an effective potential, we  need an
analytical expression for the Fourier transformation. So we have simulated
the numerical results with various function forms, finally we decide
\begin{equation}
\label{fit}
{f\over{\bf k}^{\;2}}={A\over {\bf k}^{\;2}}+{B\over {\bf k}^{\;2}+C^2},
\end{equation}
where $\bf k$ is the exchanged three-momentum, gives the best fit. The parameters
$A,B$ and $C$ are numerical values. In this expression, we keep the explicit
$k-$dependence of the form factors. In fact, the expression (\ref{fit}) can
be rewritten as
\begin{equation}
{f\over {\bf k}^{\; 2}}={(A+B)({\bf k}^{\; 2}+{A\over A+B}C^2)\over {\bf k}^{\; 2}
({\bf k}^{\; 2}+C^2)},
\end{equation}
and ${\bf k}^{\; 2}+C^2=-(k^2-C^2)$ at the case $k_0=0$.
It is the familiar pole-like form factor which is widely used in phenomenology
\cite{Anse}.

(b) We derive the effective potential by calculating the elastic scattering
amplitude, then we need to turn the corresponding quantities into the quantum mechanics
operators. The polarization vectors $\eta$ or $\eta'$ of the axial vector
diquarks must be normalized as $\eta^2=\eta^{'2}=-1$ according to the quantum field
theory. Turning $\eta (\eta')$ into QM spin-operator, we have
\begin{equation}
\eta={1\over\sqrt 2}(({\vec{\beta}}\cdot{\bf S})\gamma, {\bf S}+{\gamma-1\over{\vec{\beta}}^2}
({\vec{\beta}}\cdot{\bf S}){\vec{\beta}}),
\end{equation}
where ${\bf S}$ is the spin-operator, ${\vec{\beta}}={\bf p}/M$, $\gamma=E/M$
are the boost factor and ${\bf p}, E, M$ are the momentum, energy and mass
respectively, the factor $1/\sqrt 2$ guarantees the right normalization for
the axial-vector diquark $S(S+1)=2$.

In derivation of the potential, we first calculate the part
corresponding to the scattering amplitude induced by one-gluon
exchange,
\begin{equation}
M^{gluon}({\bf p},{\bf k})=<\lambda^a\lambda^a>
g^2_s\frac{1}{16\sqrt{E_1 E^\prime_1 E_2 E^\prime_2}}
\bar{u}({\bf p}^\prime_1)\gamma^\nu u({\bf p}_1)D_{\mu\nu}({\bf k})\langle p^\prime_2|J^\mu|
p_2\rangle,
\end{equation}
where the Coulomb gauge for the gluon propagator is chosen,
\begin{eqnarray}
D^{00}(k)=-\frac{1}{{\bf k}^2},
D^{ij}(k)=-\frac{1}{k^2}(\delta^{ij}-\frac{k^i k^j}{{\bf k}^2}),
D^{0i}=D^{i0}=0.
\end{eqnarray}
Thus we have the expressions for the transition amplitudes as following:
\begin{eqnarray}
V_{gluon}^{SS}({\bf p},{\bf k}) &=& -\frac{16\pi\alpha_s}{3}\frac{f}{{\bf k}^2}
[1+\frac{{\bf p}^2}{m_1m_2}
-\frac{{\bf k}^2}{4m_1}(\frac{1}{2m_1}+\frac{1}{m_2})\nonumber \\
&+& \frac{i{\bf S}_1\cdot({\bf k}\times{\bf p})}
{2m_1}(\frac{1}{m_1}+\frac{2}{m_2})], \; \hspace{2cm}
\;\;({\rm for\;Sq\to Sq});\\
V_{gluon}^{AA}({\bf p},{\bf k}) &=&-\frac{16\pi\alpha_s}{3}\frac{f}{{\bf{k}}^2}
[1+\frac{{\bf p}^2}{m_1m_2}
-\frac{{\bf{k}}^2}{4m_1}(\frac{1}{2m_1}+\frac{1}{m_2})
 \nonumber\\
&+& \frac{i{\bf{S}}_1\cdot({\bf{k}}\times{\bf{p}})}{2m_1}(\frac{1}{m_1}+\frac{2}{m_2})
+\frac{i{\bf{S}}_2\cdot({\bf{k}}\times{\bf{p}})}{4m_2}(\frac{1}{m_1}+\frac{1}{m_2})
\nonumber\\
&-& \frac{({\bf{S}}_1\cdot{\bf{S}}_2){\bf{k}}^2-({\bf{S}}_1
\cdot{\bf{k}})({\bf{S}}_2\cdot{\bf{k}})}{4m_1m_2}],\; \hspace{1cm}
({\rm for \; Aq\to Aq});\\
V_{gluon}^{AS}({\bf{p}},{\bf{k}}) &=&
\frac{16\pi\alpha_s}{3}\frac{f}{2\sqrt{2}{\bf{k}}^2}
[\frac{i{\bf{S}}_2({\bf{k}}\times{\bf{p}})}{m_2}(\frac{1}{m_1}+\frac{1}{m_2})
\nonumber \\
&-& \frac{({\bf{S}}_1\cdot{\bf{S}}_2){\bf{k}}^2-
({\bf{S}}_1\cdot{\bf{k}})({\bf{S}}_2\cdot{\bf{k}})}{m_1m_2}],
\;\;\;\;({\rm for \; Aq\to Sq\;or\; Sq\to Aq}).
\end{eqnarray}
The Fourier transformation would bring up the ordering problem which will be
discussed below, and then we will present the expressions in the configuration
space in subsection (e),
while the confinement part is given in subsection (d).

(c) Ordering of operators.

When we derive the scattering amplitude in the momentum-space, all
quantities are commutative, however, when we transform them into
the QM operators and carry out a Fourier transformation to the
configuration space, there exists an ordering problem in general.

When we transform $\bf k$($={\bf p}_1^{\;'}
-{\bf p}_1$) into $\bf r$,  which in fact is the relative radial vector
between the light quark and the heavy diquark,  if we choose the center
of mass of the system to be the coordinate origin, ${\bf r}_2$ of the
heavy diquark would be very close to zero and ${\bf r}_1\sim \bf r$.
The reduced mass of the system $mM/(m+M)\rightarrow m$, which is almost
the mass $m$ of the light quark. The momentum $\bf p$ would remain as
a derivative operator in the configuration space, thus the ordering
problem emerges.

For example, there are three different orders for $\hat{\bf p},\hat{\bf p},
g({\bf r})$
where $g({\bf r})$ is a function of $\bf r$, as
$$g({\bf r})\hat{{\bf p}}^{\; 2},\;\;\;\; \hat{\bf p}\cdot 
g({\bf r})\hat{\bf p},\;\;\;
{1\over 4}[g({\bf r})\hat{{\bf p}}^{\; 2}+2\hat{\bf p}\cdot 
g({\bf r})\hat{\bf p}+
\hat{{\bf p}}^{\; 2}g({\bf r})].$$

In most literatures, one just simply takes $g({\bf r})\hat{{\bf p}}^{\; 2}$. 
In our
work, we compare the different ordering schemes and our numerical results show
that different schemes would lead to different parametrizations, but the final
measurable spectra are not sensitive to the ordering at all.

(d) The confinement part.

The confinement part of the potential is fully due to the non-perturbative
QCD effects and is not derivable in any established theoretical framework.
So far, one can only postulate its form and determine the concerned parameters
by fitting data.

The most commonly adopted confinement form is the linear potential $V_{conf}
=ar+b$ at the leading order. It can be split into a scalar and vector
pieces which may lead to different relativistic corrections. Since its source
is obscure so far, one cannot decide fractions of each piece. But in general,
it can be written as \cite{Ebert}
\begin{eqnarray}
\label{conf}
V_{conf}^{S0}(r) &=& \kappa(ar+b), \\
V_{conf}^{V0}(r) &=& (1-\kappa)(ar+b).
\end{eqnarray}
The resultant potentials  with all relativistic corrections are
\begin{eqnarray}
V_{conf}^S(r)&=& V_{conf}^{S0}(r)-\frac{1}{2m^2_1}{{\bf{p}}}
\cdot V^{S0}(r){{\bf{p}}}+
\frac{1}{8m^2_1}\triangle V^{S0}(r)
\nonumber \\
&-&\frac{1}{2m^2_1}\frac{V^{\prime S0}(r)}{r}{\bf{S}}_1\cdot{\bf{L}}
-\frac{1}{2m^2_2}\frac{V^{\prime S0}(r)}{r}{\bf{S}}_2\cdot{\bf{L}},
\\
V_{conf}^V(r)&=&V_{conf}^{V0}(r)
-\frac{1}{4m_1}(\frac{1}{2\mu}+\frac{1}{2m_2}-\frac{1+\kappa}{m_1})
\triangle V^{V0}(r)
\nonumber \\
&+&\frac{1}{m_1m_2}{\bf{p}}\cdot V^{V0}(r){\bf{p}}
+\frac{1}{m_1}(\frac{1+\kappa}{\mu}-\frac{1}{2m_1})\frac{V^{\prime V0}(r)}{r}{\bf{S}}_1\cdot{\bf{L}}
\nonumber \\
&-&\frac{1}{2m_2^2}\frac{V^{\prime V0}(r)}{r}{\bf{S}}_2\cdot{\bf{L}}
+\frac{2(1+\kappa)}{3m_1m_2}\triangle V^{V0}(r){\bf{S}}_1\cdot{\bf{S}}_2
\nonumber \\
&+&\frac{1+\kappa}{3m_1m_2}(\frac{V^{\prime V0}(r)}{r}-
V^{\prime\prime V0}(r))S_{12},
\end{eqnarray}
and
$$ S_{12}\equiv {\bf S}_1\cdot{\bf S}_2-{3\over r^2}({\bf{S}}_1\cdot{\bf r})({\bf S}_2
\cdot{\bf r}).$$

In later numerical calculations, we choose several values of $\kappa$.

(e) The potential.

Finally we have the full Hamiltonian
\begin{equation}
\label{hal}
H=K+V,
\end{equation}
where $K$ is the kinetic part and
\begin{equation}
V=V_{gluon}+V_{conf},
\end{equation}
and
\begin{equation}
V_{conf}=V_{conf}^V+V_{conf}^S.
\end{equation}
The single gluon exchanged potential $V_{gluon}$ has the following forms as
\begin{eqnarray}
V_{gluon}^{SS}(r)&=&-\frac{4\alpha_s}{3}[\frac{A}{r}+\frac{A}{4m_1m_2}(\frac{1}{r}\hat{{\bf{p}}}^2
+2\hat{{\bf{p}}}\cdot \frac{1}{r}\hat{{\bf{p}}}+\hat{{\bf{p}}}^2\frac{1}{r})
\nonumber \\
&-&\frac{\pi A}{m_1}(\frac{1}{2m_1}+\frac{1}{m_2})\delta({\bf{r}})
-\frac{A}{2m_1}(\frac{1}{m_1}+\frac{2}{m_2})\frac{{\bf{S}}_1\cdot{\bf{L}}}{r^3}
\nonumber \\
&+&\frac{Be^{-Cr}}{r}+\frac{B}{4m_1m_2}(\frac{e^{-Cr}}{r}\hat{{\bf{p}}}^2
+2\hat{{\bf{p}}}\cdot \frac{e^{-Cr}}{r}\hat{{\bf{p}}}+\hat{{\bf{p}}}^2\frac{e^{-Cr}}{r})
\nonumber \\
&+&\frac{BC^2}{4m_1}(\frac{1}{2m_1}+\frac{1}{m_2})\frac{e^{-Cr}}{r}
 \nonumber \\
&-& \frac{B}{2m_1}(\frac{1}{m_1}+\frac{2}{m_2})
\frac{(Cr+1)e^{-Cr}}{r^3}{\bf{S}}_1\cdot{\bf{L}}],
\;\;\;({\rm for \; scalar-diquark+q \; baryons})    ;\\
V_{gluon}^{AA}(r)&=&-\frac{4\alpha_s}{3}[\frac{A}{r}+\frac{A}{4m_1m_2}(\frac{1}{r}\hat{{\bf{p}}}^2
+2\hat{{\bf{p}}}\cdot \frac{1}{r}\hat{{\bf{p}}}+\hat{{\bf{p}}}^2\frac{1}{r})
\nonumber \\
&-&\frac{\pi A}{m_1}(\frac{1}{2m_1}+\frac{1}{m_2})\delta({\bf{r}})
-\frac{A}{2m_1}(\frac{1}{m_1}+\frac{2}{m_2})\frac{{\bf{S}}_1\cdot{\bf{L}}}{r^3}
-\frac{A}{4m_2}(\frac{1}{m_1}+\frac{1}{m_2})\frac{{\bf{S}}_2\cdot{\bf{L}}}{r^3}
\nonumber \\
&-&\frac{A}{4m_1m_2}\frac{S_{12}}{r^3}
-\frac{2\pi A}{3m_1m_2}({\bf{S}}_1\cdot{\bf{S}}_2)\delta({\bf{r}})
\nonumber \\
&+&\frac{Be^{-Cr}}{r}+\frac{B}{4m_1m_2}(\frac{e^{-Cr}}{r}\hat{{\bf{p}}}^2
+2\hat{{\bf{p}}}\cdot \frac{e^{-Cr}}{r}\hat{{\bf{p}}}+\hat{{\bf{p}}}^2\frac{e^{-Cr}}{r})
\nonumber \\
&+&\frac{BC^2}{4m_1}(\frac{1}{2m_1}+\frac{1}{m_2})\frac{e^{-Cr}}{r}
-\frac{B}{2m_1}(\frac{1}{m_1}+\frac{2}{m_2})\frac{(Cr+1)e^{-Cr}}{r^3}{\bf{S}}_1\cdot{\bf{L}}
\nonumber \\
&-&\frac{B}{4m_2}(\frac{1}{m_1}+\frac{1}{m_2})\frac{(Cr+1)e^{-Cr}}{r^3}{\bf{S}}_2\cdot{\bf{L}}
\nonumber \\
&-&\frac{B}{12m_1m_2}\frac{C^2r^2+3Cr+3}{r^3}e^{-Cr}S_{12} \nonumber\\
&+& \frac{B}{6m_1m_2}\frac{C^2e^{-Cr}}{r}{\bf{S}}_1\cdot{\bf{S}}_2],
\;\;\; ({\rm for\; axial-vector-diquark+q\; baryons});\\
\label{SA}
V_{gluon}^{SA}(r)&=&-\frac{4\alpha_s}{3}
[\frac{A}{2\sqrt{2}m_2}(\frac{1}{m_1}+\frac{1}{m_2})\frac{{\bf{S}}_2\cdot{\bf{L}}}{r^3}
+\frac{A}{2\sqrt{2}m_1m_2}\frac{S_{12}}{r^3}
+\frac{4\pi A}{3\sqrt{2}m_1m_2}({\bf{S}}_1\cdot{\bf{S}}_2)\delta({\bf{r}})
\nonumber \\
&+&\frac{B}{2\sqrt{2}m_2}(\frac{1}{m_1}+\frac{1}{m_2})\frac{(Cr+1)e^{-Cr}}{r^3}{\bf{S}}_2\cdot{\bf{L}}
\nonumber \\
&+&\frac{B}{6\sqrt{2}m_1m_2}\frac{C^2r^2+3Cr+3}{r^3}e^{-Cr}S_{12} \nonumber \\
&-& \frac{B}{3\sqrt{2}m_1m_2}\frac{C^2e^{-Cr}}{r}{\bf{S}}_1\cdot{\bf{S}}_2],
\;\;\;\;({\rm for\; mixing\; between\; scalar-diquark+q}\nonumber \\
&&{\rm and\; axial-vector-diquark+q\; baryons)},
\end{eqnarray}
where
\begin{eqnarray}
S_{12}=\frac{3({\bf{S}}_1\cdot{\bf{r}})({\bf{S}}_2\cdot{\bf{r}})}{r^2}
-{\bf{S}}_1\cdot{\bf{S}}_2.
\end{eqnarray}
In the expressions $A$ and $S$ stand for the axial-vector and scalar respectively.
Later we will show that even though $V_{gluon}^{SA}$ derived in QFT is not
trivially zero, in the non-relativistic QM framework, it can give only null
contribution. We will discuss this issue in the last section.\\

\noindent{\bf III. The variational method}

(a) We choose the variational trial function with a single parameter for the
1S state as
\begin{equation}
\label{tri}
\psi(r,\theta,\phi)=R(r)Y_{00}(\theta,\phi),
\end{equation}
and
\begin{equation}
R(r)=Ne^{-\lambda r^{\delta}},
\end{equation}
where $\delta=4/3$ and the normalization is
$$N= \left( {\delta (2\lambda)^{3/\delta}\over\Gamma({3\over\delta})}\right)^{1/2}.
$$
and $\lambda$ is the variational parameter. This form is  discussed in our
earlier work \cite{Ding} where we tested some $\delta-$values and found that
$\delta=4/3$ would be more appropriate for the Cornell-type potential, we will
discuss this problem further in the last section.

(b) The parameters $\alpha_s, a$ and $b$.

As noticed, the relativistic effects are serious because of the existence of a
light quark. Unlike the heavy quarkonium, such as $J/\psi,\Upsilon$ etc.,
truncation of the non-relativistic expansion where we only keep it up to
${\bf p}^{\; 2}/m^2$ order, is not a good approximation. However, we can partly
compensate the effects by attributing the uncertainties to the
potential parameters which are not directly measurable. In other words, in the
process of fitting data of mesons containing a light quark, such as $B^{(*)}$,
we have attributed the unknown factors into
the phenomenological parameters, then later when we use the set of parameters
to evaluate the spectra of baryons containing two heavy quarks and a light quark,
the non-perturbative QCD effects and the relativistic influence are or at least
mostly included. Obviously in the case, if one used the parameters obtained by
fitting data of heavy quarkonia, the errors are un-controllable. Here we choose
$B^{(*)}$ data to obtain $\alpha_s, a$ and $b$. It is worth noticing that
in the D-case, the relativistic effects are serious and the charm-quark  is not
heavy enough, so when we apply our trial function to the D-case, we find
the minimum of the expectation value of energy is not stable. Thus we abandon
the D-case. When we use the variational method to obtain the parameters, we
retain all the relativistic corrections in the potential for $B^{(*)}-$mesons.

(c) Then we turn to calculate the spectra of the baryons containing two
heavy quarks, thus $\lambda$ stands as the variational parameter. The
expectation value of $H$ is
\begin{equation}
E(\lambda)=<H>={<R(\lambda)|H|R(\lambda)>\over <R(\lambda)|R(\lambda)>},
\end{equation}
where $R(\lambda)$ is the chosen trial function (\ref{tri}). Then minimizing
$E(\lambda)$ as
$${dE(\lambda)\over d\lambda}=0,$$
we obtain the $\lambda-$value. In the expression $H$ is the full Hamiltonian
given in eq.(\ref{hal}).

The advantage of using the variational method is obvious, that is
we are able of
treating all terms simultaneously. Unlike the perturbation method where all
relativistic corrections which are very large in this case are dealt with
perturbatively, so that remarkable errors for the baryons which contains not
only two heavy quarks but also a light one, emerge due to the ill-treatments,
by contraries, the ambiguities can be avoided in our treatments.\\

\noindent{\bf IV. The numerical results}

We are listing some concerned parameters which appear in our formulae.

The $A,B,C-$values in eq.(\ref{fit}) are
$$A=1.0060428,\;\;\; B=-1.280532,\;\;\; C=2.425356, \;\;\;\;{\rm for\; cc\;
diquark};$$
$$A=1.0019211,\;\;\; B=-1.545131,\;\;\; C=6.415003, \;\;\;\;{\rm for\; bc\;
diquark};$$
$$A=1.0005426,\;\;\; B=-1.244000,\;\;\; C=3.676031, \;\;\;\;{\rm for\; bb\;
diquark}.$$
We have the constituent quark masses and the heavy diquark masses as
$$m_u=m_d=0.33\;GeV,\; m_s=0.5\; GeV,\;M_{cc}=3.26\;GeV,\; M_{bc}=6.52\;GeV,
\;M_{bb}=9.79\;GeV.$$
It is noted that $bb$ and $cc$ diquark must be axial vectors, but $bc$ can
be either a scalar or an axial vector, the mass splitting of the
scalar and axial-vector $bc$ diquarks can be neglected in practical
calculations.

The baryon spectra are calculated and the results are given in Tables 1 and 2,
with q being u or d.
In Table 1, we choose $\kappa=-1$ for the confinement potential (\ref{conf})
which is consistent with that used in ref.\cite{Ebert}, and list results
corresponding to various ordering schemes. In Table 2, we change the
$\kappa-$values in the confinement potential and use the ordering scheme 2, i.e. $\hat{\bf p}\cdot g(\bf r)\hat{\bf p}$.

\begin{center}
Table 1.\\
\vspace{0.3cm}
\begin{tabular}{|c|c|c|c|}
\hline
type & Ordering 1 & Ordering 2 & Ordering 3 \\
\hline
(ccq)(1/2) & e=0.1972, $M_B$=3.787 & e=0.2175, $M_B$=3.808 & e=0.1480,
$M_B$=3.738\\
\hline
(ccq)(3/2) & e=0.2229, $M_B$=3.813 & e=0.2425, $M_B$=3.832 & e=0.1793, $M_B$=3.772\\
\hline
(ccs)(1/2) & e=0.2027, $M_B$=3.963 & e=0.1101, $M_B$=3.870 & e=0.0875, $M_B$=3.851\\
\hline
(ccs)(3/2) & e=0.2254, $M_B$=3.985 & e=0.1375, $M_B$=3.897 & e=0.1187, $M_B$=3.879\\
\hline
(cbq)(1/2)$_S$ & e=0.2411, $M_B$=7.091 & e=0.2325, $M_B$=7.082 & e=0.1931, $M_B$=7.043\\
\hline
(cbq)(1/2)$_A$ & e=0.2320, $M_B$=7.082 & e=0.2234, $M_B$=7.073 & e=0.1823, $M_B$=7.032\\
\hline
(cbq)(3/2)$_A$ & e=0.2455, $M_B$=7.095 & e=0.2369, $M_B$=7.087 & e=0.1981, $M_B$=7.048\\
\hline
(cbs)(1/2)$_S$ & e=0.2268, $M_B$=7.247 & e=0.1202, $M_B$=7.140 & e=0.1158, $M_B$=7.236\\
\hline
(cbs)(1/2)$_A$ & e=0.2181, $M_B$=7.238 & e=0.1096, $M_B$=7.130 & e=0.1073, $M_B$=7.225\\
\hline
(cbs)(3/2)$_A$ & e=0.2310, $M_B$=7.251 & e=0.1253, $M_B$=7.145 & e=0.1210, $M_B$=7.241\\
\hline
(bbq)(1/2) & e=0.2168, $M_B$=10.337 & e=0.1983, $M_B$=10.318 & e=0.1731, $M_B$=10.293\\
\hline
(bbq)(3/2) & e=0.2251, $M_B$=10.345 & e=0.2069, $M_B$=10.327 & e=0.1830, $M_B$=10.303\\
\hline
(bbs)(1/2) & e=0.2096, $M_B$=10.500 & e=0.0910, $M_B$=10.381 & e=0.1061, $M_B$=10.396\\
\hline
(bbs)(3/2) & e=0.2174, $M_B$=10.507 & e=0.1009, $M_B$=10.390 & e=0.1150, $M_B$=10.405\\
\hline
\end{tabular}
\end{center}

\vspace{0.3cm}
In Table 1, "ordering 1" means $g({\bf r})\hat{\bf p}^{\; 2}$, where $\alpha_s=0.23,
a=0.11,b=-0.13$; "ordering 2" means $\hat{\bf p}\cdot 
g({\bf r})\hat{\bf p}$ with
$\alpha_s=0.41,a=0.09,b=-0.21$; "ordering 3" means $(g({\bf r})\hat{\bf p}^{\; 2}
+2\hat{\bf p}\cdot 
g({\bf r})\hat{\bf p}+\hat{\bf p}^{\; 2}g({\bf r}))/4$ with $\alpha_s=0.23,
a=0.11,b=-0.27$. The subscript A and S stand for the axial vector and scalar
respectively. $e$ is the binding energy and $M_B$ is the baryon mass with unit
GeV. In the calculations, $\kappa=-1.$\\

\begin{center}
Table 2\\
\vspace{0.3cm}

\begin{tabular}{|c|c|c|c|}
\hline
Type & $\kappa=0$ & $\kappa=0.5$ & $\kappa=1.0$ \\
\hline
(ccq)(1/2) & e=0.2133, $M_B$=3.703 & e=0.2265, $M_B$=3.817 & e=0.1713, $M_B$=3.761\\
\hline
(ccq)(3/2) & e=0.2449, $M_B$=3.735 & e=0.2583, $M_B$=3.848 & e=0.2153, $M_B$=3.805\\
\hline
(ccs)(1/2) & e=0.1466, $M_B$=3.807 & e=0.0834, $M_B$=3.843 & e=0.0501, $M_B$=3.810\\
\hline
(ccs)(3/2) & e=0.1756, $M_B$=3.836 & e=0.1180, $M_B$=3.878 & e=0.0960, $M_B$=3.856\\
\hline
(cbq)(1/2)$_S$ & e=0.2539, $M_B$=7.104 & e=0.2582, $M_B$=7.108 & e=0.2294, $M_B$=7.079\\
\hline
(cbq)(1/2)$_A$ & e=0.2428, $M_B$=7.093 & e=0.2463, $M_B$=7.096 & e=0.2142, $M_B$=7.064\\
\hline
(cbq)(3/2)$_A$ & e=0.2593, $M_B$=7.109 & e=0.2640, $M_B$=7.114 & e=0.2368, $M_B$=7.087\\
\hline
(cbs)(1/2)$_S$ & e=0.1799, $M_B$=7.200 & e=0.1045, $M_B$=7.125 & e=0.1048, $M_B$=7.125\\
\hline
(cbs)(1/2)$_A$ & e=0.1693, $M_B$=7.189 & e=0.0897, $M_B$=7.110 & e=0.0885, $M_B$=7.108\\
\hline
(cbs)(3/2)$_A$ & e=0.1851, $M_B$=7.205 & e=0.1116, $M_B$=7.131 & e=0.1126, $M_B$=7.133\\
\hline
(bbq)(1/2) & e=0.2131, $M_B$=10.333 & e=0.2187, $M_B$=10.339 & e=0.1748, $M_B$=10.295\\
\hline
(bbq)(3/2) & e=0.2236, $M_B$=10.344 & e=0.2298, $M_B$=10.350 & e=0.1894, $M_B$=10.309\\
\hline
(bbs)(1/2) & e=0.0876, $M_B$=10.378 & e=0.0677, $M_B$=10.357 & e=0.0607, $M_B$=10.351\\
\hline
(bbs)(3/2) & e=0.0997, $M_B$=10.390 & e=0.0813, $M_B$=10.371 & e=0.0759, $M_B$=10.366\\
\hline
\end{tabular}
\end{center}

\vspace{0.3cm}

In Table 2, for the confinement potential $V_{conf}^S=\kappa(ar+b),
V_{conf}^A=(1-\kappa)(ar+b)$, we have:
for $\kappa=0$, the fitted $\alpha_s=0.44,a=0.14,b=-0.37$; for
$\kappa=0.5$,  the fitted $\alpha_s=0.5, a=0.14,b=-0.37$; for $\kappa=1.$
the fitted $\alpha_s=0.73,
a=0.16,b=0.45$. Here we use the ordering 2, i.e. $\hat{\bf p}\cdot
g({\bf r})\hat{\bf p}$.

\vspace{0.5cm}

\noindent{\bf V. Conclusion and discussion}

As we noted in the introduction, we calculate the spectra of baryons
containing two heavy quarks which can constitute a  diquark. It is believed
that such a subject can be spatially tight and serve as a color source for the
light quark. This picture greatly simplifies the calculations. The difference
of our method from previous works is in several aspects.

First, we use effective vertices
$DD'g$ where $D$ and $D'$ are scalar or axial-vector diquarks and $g$ is gluon.
We derive the form factors at the vertices based on the B-S equation and so we
can keep their explicit $k^2-$dependence which leads to an extra Yukawa-type
term in the potential.

Secondly, we investigate the ordering problem which is brought up by the
Fourier transformation with respect to the exchanged momentum $\bf k$ and the
quantization of momentum $\bf p$. We find that various ordering schemes
can lead to different parametrizations for $\alpha_s$ (note that here
$\alpha_s$ is an effective coupling constant, but not that from the fundamental
QCD theory), and $a,b$ which are not directly measurable. However, we find
that the final
results do not deviate much from each other. So we can conclude that at least
the ordering schemes do not seriously influence the spectra evaluation in
the variational method. In our future work, we will continue to investigate
if the ordering schemes can induce other observable effects such as the
effective decay constant etc.

In our scenario, we use the variational method and the Hamiltonian includes
not only the leading kinetic and potential terms, but also the relativistic
corrections up to order of ${\bf p}^{\; 2}$. Because the relativistic effects
are very serious in the case where a light flavor is involved, this treatment
is superior to the perturbative method. To reduce the uncertainties and
errors brought up by the truncation of the non-relativistic expansion, we
use the $B^{(*)}$ data where a light quark is moving around the heavy
b-quark, as inputs to obtain suitable parametrization. As hoped, most of
those uncertainties and errors can be attributed into the phenomenological
parameters $\alpha_s,a$ and $b$.

As well-known, the potential model cannot perfectly describe hadron
characteristics which are mostly determined by non-perturbative QCD effects and we have
no reliable knowledge on it so far. But as long as we use the experimental
data as inputs to parametrize the model, the disadvantages can be partly
compensated. In the trial function $R(\lambda,r)$ in the form of eq.(\ref{tri}),
we priori take $\delta=4/3$
based on our previous work. In fact, $\delta$ should be an irrational value
between 1 and 2 which correspond to the solutions for the Coulomb and harmonic
oscillator potentials. But $\delta=4/3$, as indicated in ref.\cite{Ding}, is
a satisfactory value for the Cornell potential. Numerically, in our
previous work\cite{Ding}, we obtained
the irrational number $\delta$ for the Cornell potential as $\delta=1.33809...$, which is very close to 4/3. Then we take the value and only
let $\lambda$ be the unique variational parameter, this treatment can greatly
simplify the calculations with sufficient accuracy being kept.

Even though the diquark picture is believed to work in this case and the derived form
factors further improves the situation, there still exists small deviation
from reality, including the diquark masses. This should be further investigated.

Finally, as we pointed above, although the mixing term in eq.(\ref{SA})
which is
derived in QFT is not trivially zero, when we sandwich it among the quantum
states, we have
\begin{equation}
<\psi(1/2,A,l=0)|V_{gluon}^{(SA)}|\psi(1/2,S,l=0)>
=<\psi(1/2,A,l=1)|V_{gluon}^{(SA)}
|\psi(1/2,S,l=1)>\equiv 0.
\end{equation}
The matrix elements are absolutely zero. The reason is simple, because in the
framework of non-relativistic quantum mechanics, there are no creation and
annihilation operators as in  QFT, so that we can only deal with elastic
scattering. The mixing between $\psi(1/2,A)$ and $\psi(1/2,S)$ refer to a
change of spin or particle identity, so cannot appear in QM even though we know
such mixing must exist and may play important roles to hadron spectra. For
example as in a completely different area of the hadron spectroscopy,
the mixing between glueball and quarkonium is known as very important or
even crucial to phenomenology, but we cannot evaluate it in the
potential model. We will further study these mixing effects in our future
work \cite{Li}.

The B-factory and other facilities of high energy experiments may provide data
on $\Xi_{cc}^{(*)}$ and other such baryons. Once the data are available, we may
re-adjust our input parameters and make further predictions on the spectra and
other characters of the baryons, then we can testify the validity of the diquark picture
and the non-relativistic potential model.\\

\noindent{\bf Acknowledgments:}

This work is partially supported by the National Natural Science Foundation
of China and the Australian Research Council.
This work is also part of Tong Sheng-Ping's Master thesis. One of the
authors(Ding) would like to thank Prof. G. Prosperi for his hospitality
during Ding's visit in the Department of Physics, Univ. of Milan.

\end{document}